\documentclass[]{article}

\begin{document}

\title{Chisholm-Caianiello-Fubini Identities for $S=1$
Barut-Muzinich-Williams Matrices\thanks{This the modernized version of 
a EF-UAZ FT-95-14 unpublished preprint of 1995 of the second author. The modifications are due 
to the Thesis of the first author.}}

\author{{\bf M. de G. Caldera Cabral} and {\bf V. V. Dvoeglazov}\\
UAF, Universidad Aut\'onoma de Zacatecas \\
Apartado Postal 636, Suc. 3 Cruces, Zacatecas 98068, Zac., M\'exico\\
E-mail:  valeri@fisica.uaz.edu.mx}

\date{}

\maketitle

\begin{abstract}
The formulae of the relativistic  products 
are found $S=1$ Barut-Muzinich-Williams matrices. They are analogs of the well-known Chisholm-Caianiello-Fubini
identities. The obtained results can be useful in
the higher-order calculations of the high-energy processes 
with $S=1$ particles  in the framework of the $2(2S+1)$ Weinberg formalism, 
which recently attracted attention again.

\end{abstract}



\newpage

The attractive Weinberg $2(2S+1)$ component formalism for description
of higher spin particles~\cite{Weinberg} is based on the same principles
as the Dirac formalism for spin-1/2, Ref.~\cite{Ryder}. Further 
developments~\cite{Greiner,Dvoeglazov-old,Ahluwalia,Dvoeglazov,Dvoeglazov2}
showed that many interesting things can be found therein. For instance,
the connections with the modified Bargmann-Wigner formalism~\cite{Dvoeglazov2}
or the connections with the so-called Bargmann-Wightman-Wigner 
formalism (BWW)~\cite{Gelfand,BWW,Ahluwalia,Dvoeglazov}.
On the basis of the analysis of
the $(S,0)\oplus (0,S)$ representation space it was found there that
the intrinsic parities of boson and its antiboson can be opposite, see
also~\cite{Silagadze}.  ``If a neutrino is
identified with the self/anti-self charge-conjugate representation space,
then it may be coupled with the BWW bosons to generate physics beyond
the present day gauge theories.", see the above-cited references.
One more hint at the possible future application of these formalisms is the
tentative experimental evidence for a tensor coupling in the $\pi^- \rightarrow
e^- + \tilde\nu_e +\gamma$ decay, for instance~\cite{Bolotov}.
There exist  experimental opportunities to check the
existence of the ``unconventional"  bosons and fermions, and different types 
of interactions as well, beyond the Standard Model, e.~g., Ref.~\cite{DME}.

The principal equation in this formalism is that of the ``$2s$"-order in the momentum
operators. The analogs of the Dirac $\gamma$-matrices have also ``$2s$"
vectorial indices:
\begin{equation}
[\gamma_{\mu_1 \mu_2...\mu_{2s}} \partial^{\mu_1}\partial^{\mu_2}...\partial^{\mu_2s} + m^{2s}]
\Psi (x) = 0.
\end{equation}
The covariant-defined $\Gamma$-~matrices for any spin have been
introduced by Barut, Muzinich and Williams~\cite{Barut}, see
also~\cite{Shay,TH}.  For the case of spin $S=1$ they have the following
form:\footnote{The Eiclidean metric is used.}
\begin{eqnarray}
&&\Gamma^{(1)} \equiv I = \pmatrix{ I & 0\cr 0 &
I} ,\nonumber\\
&&\Gamma^{(2)} \equiv \gamma_5 = \pmatrix { - I & 0\cr 0 &
I} , \nonumber\\
&&\Gamma^{(3)}_{\alpha\beta} \equiv \gamma_{\alpha\beta} = \pmatrix
{0 & - \tilde S_{\alpha\beta}^{\dagger}\cr - \tilde S_{\alpha\beta} & 0} , \nonumber\\
&&\Gamma^{(4)}_{\alpha\beta} \equiv \gamma_{4, \alpha\beta} = 
i\gamma_5 \gamma_{\alpha\beta} , \\
&&\Gamma^{(5)}_{\alpha\beta} \equiv \gamma_{5, \alpha\beta}
= i \left [\gamma_{\alpha\lambda},
\gamma_{\beta\lambda}\right ]_{-} , \nonumber\\
&&\Gamma^{(6)}_{\alpha\beta, \mu\nu} \equiv \gamma_{6, \alpha\beta, \mu\nu}
=
 \left [\gamma_{\alpha\mu}, \gamma_{\beta\nu} \right ]_{+}
+2\delta_{\alpha\mu} \delta_{\beta\nu} - \left [\gamma_{\alpha\nu},
\gamma_{\beta\mu}\right ]_{+} -2\delta_{\alpha\nu}\delta_{\beta\mu}= \nonumber\\
&&= - {1\over 12} \left [\gamma_{5, \alpha\beta}, \gamma_{5, \mu\nu}\right ]_{+}
+ 4\left (\delta_{\alpha\mu}\delta_{\beta\nu} -\delta_{\alpha\nu} \delta_{\beta\mu} \right )
- 4\epsilon_{\alpha\beta\mu\nu} \gamma_5 \,, \nonumber
\end{eqnarray}
where
\begin{eqnarray}
\tilde S_{44} &=& -I, \quad \tilde S_{i4}=\tilde S_{4i} = iS_i , \nonumber\\
\tilde S_{ij} &=& S_{ij} - \delta_{ij} = S_i S_j + S_j S_i - \delta_{ij} \,.
\end{eqnarray}
$S_i$ are the spin-1 matrices, and $\epsilon_{1234} = 1$.
They have the simmetry properties~\cite{Shay}:
\begin{eqnarray}
&&\gamma_{\alpha\beta} = \gamma_{\beta\alpha},\quad \sum_\alpha\,
\gamma_{\alpha\alpha} =0 , \nonumber\\
&&\gamma_{4, \alpha\beta} = \gamma_{4, \beta\alpha},\quad \sum_\alpha\,
\gamma_{4, \alpha\alpha} =0, \nonumber\\
&&\gamma_{5, \alpha\beta} = - \gamma_{5, \beta\alpha} , \\
&&\gamma_{6, \alpha\beta, \mu\nu} = - \gamma_{6, \beta\alpha, \mu\nu},\quad
\gamma_{6, \alpha\beta, \mu\nu} = \gamma_{6, \mu\nu, \alpha\beta} , \nonumber\\
&&\gamma_{6, \alpha\beta, \mu\nu} + \gamma_{6, \alpha\mu, \nu\beta}
+  \gamma_{6, \alpha\nu, \mu \beta} = 0 .\nonumber
\end{eqnarray}
The relativistic perturbation calculations of the processes
including the $S=1$ bosons will require the development technical methods
analogous to those which have been elaborated for the fermion-fermion
interaction, namely, reducing contracted products of the corresponding
$\Gamma$ matrices~\cite{Chisholm,Caianello,Good,Chisholm2}.
Our aim with this paper is to find the formulae of the relativistic
scalar products like that $\gamma_{\mu\alpha}\ldots \gamma_{\beta\mu}$.

The following relations can be deduced by straightforward
calculations:\footnote{We have also used the Wolfram MATEMATICA programm to check them.}
\begin{equation}\label{eq:rel1}
\gamma_{\mu\alpha}\gamma_{\beta\mu} =3\delta_{\alpha\beta}
-{i\over 2} \gamma_{5,\alpha\beta} ,
\end{equation}

\begin{equation}
\gamma_{\mu\alpha}\gamma_5 \gamma_{\beta\mu} = -3\gamma_5 \delta_{\alpha\beta}
-{i\over 4} \epsilon_{\alpha\beta\sigma\tau} \gamma_{5,\sigma\tau} ,
\end{equation}

\begin{eqnarray}
\lefteqn{\gamma_{\mu\alpha}\gamma_{\sigma\tau}\gamma_{\beta\mu} =
2\gamma_{\sigma\tau}\delta_{\alpha\beta} +\gamma_{\alpha\beta} \delta_{\sigma\tau} -
\gamma_{\alpha\sigma}\delta_{\tau\beta} -\gamma_{\alpha\tau}\delta_{\sigma\beta} -}\nonumber\\
&-& \gamma_{\beta\sigma}\delta_{\alpha\tau} - \gamma_{\beta\tau} \delta_{\alpha\sigma} -
i\epsilon_{\alpha\beta\sigma\mu} \gamma_{4,\tau\mu} - i\epsilon_{\alpha\beta\tau\mu}
\gamma_{4,\sigma\mu} ,
\end{eqnarray}

\begin{eqnarray}
\lefteqn{\gamma_{\mu\alpha}\gamma_{4,\sigma\tau}\gamma_{\beta\mu} =
-2\gamma_{4,\sigma\tau}\delta_{\alpha\beta} -\gamma_{4,\alpha\beta} \delta_{\sigma\tau} +
\gamma_{4,\alpha\sigma}\delta_{\tau\beta} +\gamma_{4,\alpha\tau}\delta_{\sigma\beta} +} \nonumber\\
&+& \gamma_{4,\beta\sigma}\delta_{\alpha\tau} + \gamma_{4,\beta\tau} \delta_{\alpha\sigma} -
i\epsilon_{\alpha\beta\sigma\mu} \gamma_{\tau\mu} - i\epsilon_{\alpha\beta\tau\mu}
\gamma_{\sigma\mu} ,
\end{eqnarray}

\begin{eqnarray}
\lefteqn{\gamma_{\mu\alpha} \gamma_{5,\sigma\tau} \gamma_{\beta\mu} =
2\gamma_{5,\sigma\tau}\delta_{\alpha\beta} + 2\gamma_{5,\alpha\sigma} \delta_{\beta\tau} +
2\gamma_{5,\tau\beta} \delta_{\alpha\sigma}-} \nonumber\\
&-& 2\gamma_{5,\sigma\beta} \delta_{\alpha\tau} - 2\gamma_{5,\alpha\tau} 
\delta_{\sigma\beta} +12i \left ( \delta_{\alpha\sigma}\delta_{\tau\beta} - 
\delta_{\alpha\tau} \delta_{\sigma\beta}\right  )+ \\
&+& 12i \epsilon_{\alpha\sigma\tau\beta} \gamma_{5} , \nonumber
\end{eqnarray}

\begin{equation}\label{eq:rellas}
\gamma_{\mu\alpha} \gamma_{6, \sigma\tau , \rho\phi} \gamma_{\beta\mu} = 0 .
\end{equation}
The formulae for the $S=1$ matrices which have been used above
are presented in Appendix.

\setcounter{equation}{0}

\section*{Appendix}
\appendix

Here we present the set of algebraic relations for $S=1$ spin
matrices, cf.~\cite{Varshalovich,Weaver}. We imply a summation on the repeated indices.
\begin{eqnarray}
&& S_k S_i S_k = S_i ,\\
&& S_k S_i S_j S_k = 2\delta_{ij} - S_j S_i ,\\
&& S_k S_i S_j S_l S_k = S_l S_i S_j + S_j S_l S_i - S_j \delta_{il} ,\\
&& S_k S_i S_j S_l S_m S_k =
\delta_{ij}\delta_{lm} + \delta_{im}\delta_{jl} - S_m S_l S_j S_i ,
\end{eqnarray}
and
\begin{eqnarray}
&& S_{ij} S_k = \delta_{ij} S_k +{1\over 2} \delta_{jk} S_i +{1\over 2}
\delta_{ik} S_j +{i\over 2} \left ( \epsilon_{ikl} S_{jl} + \epsilon_{jkl}
S_{il} \right ) ,\\
&& S_{ik} S_{jl} + S_{jl} S_{ik} = 2\delta_{ik} S_{jl}
+2\delta_{jl} S_{ik}+\left ( \epsilon_{ilm} \epsilon_{jkn}
-\epsilon_{ijm} \epsilon_{kln}\right ) S_{mn} ,\\
&& S_l S_{ij} S_m =
2\delta_{ij} \delta_{lm} -\delta_{im}\delta_{jl} -\delta_{jm}\delta_{il}
-\delta_{lm} S_{ij} +\nonumber\\
&& \qquad +\delta_{im}S_l S_j +\delta_{jm} S_l S_i +\delta_{il} S_j S_m +\delta_{jl} S_i S_m ,\\
&& S_l S_{ij} S_m - S_m S_{ij} S_l = \delta_{il} (S_j S_m - S_m S_j ) + \delta_{jl} (S_i S_m - S_m S_i )+\nonumber\\
&& \qquad + \delta_{im} ( S_l S_j - S_j S_l) +\delta_{jm} (S_l S_i - S_i S_l) ,\\
&& {\mbox or}\qquad \qquad \qquad \quad = -i\epsilon_{ilm} S_j - i\epsilon_{jlm} S_i
- 2\delta_{ij} (S_m S_l - S_l S_m) ,\\
&& S_i S_j S_k  + S_j S_k S_i + S_k S_i S_j = S_i \delta_{jk} + S_k \delta_{ij}
+ S_j \delta_{ik} +\nonumber\\
&& \qquad + {i\over 4} \left ( \epsilon_{ijl} S_{lk} +\epsilon_{kil} S_{jl}
+\epsilon_{jkl} S_{il}\right ) .
\end{eqnarray}
This set supplies the known formulae for $S=1$ spin matrices, e.~g.
presented in~\cite{Weaver}:
\begin{eqnarray}
&& S_i S_j S_k + S_k S_j S_i = \delta_{ij} S_k + \delta_{jk} S_i , \\
&& \tilde S_{ik} S_j =-{i\over 2} \left [ \delta_{ij} \tilde S_{4k}
+\delta_{jk} \tilde S_{4i} +\epsilon_{jil} \tilde S_{lk} +\epsilon_{jkl}
\tilde S_{il}\right ]\\
&&\Sigma_i^2 = S_3^2 ,\, \, {\mbox no \,\, summation},\\
&& \Sigma_i \Sigma_j + \Sigma_j \Sigma_i = 2\delta_{ij} S_3^2 , \\
&& \Sigma_i \Sigma_j - \Sigma_j \Sigma_i = 2i\epsilon_{ijk} \Sigma_k ,
\end{eqnarray}
where
\begin{equation}
\Sigma_1 \equiv S_1^2 -S_2^2 ,  \quad \Sigma_2 \equiv {\tilde S}_{12} = S_1 S_2 + S_2 S_1 ,
\quad \Sigma_3 \equiv S_3 .
\end{equation}


\begin{thebibliography}{99}
\footnotesize{
\bibitem{Weinberg} S. Weinberg, {\it Phys. Rev.} {\bf 133}, B1318 (1964).

\bibitem{Ryder} L. H. Ryder, {\it Quantum Field Theory.} (Cambridge Univ. Press, 1985).

\bibitem{Greiner} W. Greiner, {\it Relativistic Quantum Mechanics.} (Springer-Verlag,
Berlin-Heidelberg, 1990).

\bibitem{Dvoeglazov-old} V. V. Dvoeglazov and N. B. Skachkov, {\it Yad. Fiz.}
{\bf 48}, 1770 [English translation: {\it Sov. J. Nucl. Phys.},
1065] (1988);  JINR Communications R2-87-882 (1987); V. V. Dvoeglazov and S. V. Khudyakov,
{\it Hadronic J.} {\bf  21}, 507 (1998); V. V. Dvoeglazov , {\it Rev. Mex. Fis. Suppl.}
{\bf 40}, 352 (1994), hep-th/9401043.

\bibitem{Ahluwalia} D. V. Ahluwalia, M. B. Johnson and
T.  Goldman, {\it Phys.  Lett.} {\bf B316}, 102 (1993).

\bibitem{Dvoeglazov} V. V. Dvoeglazov, {\it Int. J. Theor. Phys.} 
{\bf 37}, 1915 (1998).

\bibitem{Dvoeglazov2} V. V. Dvoeglazov, {\it Int. J. Mod. Phys.} {\bf B20}, 1317 (2006);
{\it J. Phys. CS} {\bf 91}, 012009 (2007); {\it  Int. J. Mod. Phys. CS} {\bf 03}, 121 (2011).

\bibitem{Gelfand} I. M. Gelfand and M. L. Tsetlin, ZhETF {\bf 31}, 1107 (1956); 
G. A. Sokolik, ZhETF {\bf 33}, 1515 (1957).

\bibitem{BWW} E. P. Wigner, in {\it Group Theoretical Concepts
and Methods in Elementary Particle Physics -- Lectures of the Istanbul
Summer School of Theoretical Physics, 1962.} Ed. by F. G\"ursey (Gordon and Breach, New York-London-Paris, 1964).

\bibitem{Silagadze} Z. K. Silagadze, {\it Yad. Fiz.} {\bf 55}, 707
[English translation: {\it Sov. J. Nucl. Phys.}, 392] (1992).

\bibitem{Bolotov} V. N. Bolotov {\it et al.}, {\it Phys. Lett.}
{\bf B243}, 308 (1990).

\bibitem{DME} D. N. Casta\~no, {\it Dark Matter Constrains from High Energy Astrophysical Observations.} Ph. D. Thesis (2012), http://eprints.ucm.es/15300/1/T33772.pdf . 

\bibitem{Barut} A. Barut, I. Muzinich and D. N. Williams,
{\it Phys. Rev.} {\bf 130}, 442 (1963).

\bibitem{Shay} A. Sankaranarayanan and R. H. Good, {\it Nuovo Cimento}
{\bf 36}, 1303 (1965); D. Shay and R. H. Good, Jr., {\it Phys. Rev.}
{\bf 179}, 1410 (1969).

\bibitem{TH} R. H. Tucker and C. L. Hammer, Phys. Rev. {\bf D3}, 2448 (1971).

\bibitem{Chisholm} J. S. R. Chisholm, {\it Proc. Cambridge
Phil. Soc.} {\bf 48}, 300 (1952).

\bibitem{Caianello} E. R. Caianiello and S. Fubini, {\it Nuovo Cimento}
{\bf 9}, 1218 (1952).

\bibitem{Good} R. H. Good, Jr., {\it Rev. Mod. Phys.} {\bf 27}, 187 (1955).

\bibitem{Chisholm2} J. S. R. Chisholm, {\it Nuovo Cimento} {\bf 30}, 426
(1963); J. Kahane, {\it J. Math. Phys.} {\bf 9}, 1732 (1968);
J. S. R. Chisholm, {\it Comp. Phys. Comm.} {\bf 4}, 205 (1972).


\bibitem{Varshalovich} D. A. Varshalovich {\it et al.}, {\it Quantum Theory of
Angular Momentum.} (World Scientific, Singapore, 1988).

\bibitem{Weaver} D. L. Weaver, {\it Am. J. Phys.} {\bf 46}, 721 (1978); {\it idem.}, {\it J. Math. Phys.} {\bf 19}, 88 (1978).

}



\end{thebibliography}
\end{document}